\documentclass[oldversion]{aa}
\usepackage{graphicx}
\usepackage{txfonts}
\usepackage{lscape}
\usepackage{ulem}

\begin{document}
   \title{Secular changes in the quiescence of WZ\,Sge:\\ the development of a cavity in the inner disk}

   \author{E. Kuulkers\inst{1}
	  \and
          A.A. Henden\inst{2}
	  \and
	  R.K. Honeycutt\inst{3}
	  \and
	  W. Skidmore\inst{4}
	  \and
	  E.O. Waagen\inst{2}
          \and 
	  G.A. Wynn\inst{5}
	  }

   \authorrunning{E.~Kuulkers et al.}
   \titlerunning{Secular changes in the quiescence of WZ\,Sge}

   \offprints{E. Kuulkers}

   \institute{ISOC, ESA, European Space Astronomy Centre (ESAC), P.O.~Box 78, 28691, Villanueva de la Ca\~nada (Madrid), Spain\\
              \email{Erik.Kuulkers@esa.int}
          \and
             American Association of Variable Star Observers, 49 Bay State Rd., Cambridge, MA 02138, USA 
          \and
	     Astronomy Department, Indiana University, Swain Hall West 319, 727 East 3rd Street, Bloomington, IN 47405-7105, USA
          \and
	     TMT Observatory Corporation, 2632 E Washington Blvd, Pasadena, CA 91107, USA
          \and
	     Department of Physics and Astronomy, University of Leicester, Leicester, LE1 7RH, United Kingdom
             }

   \date{Received; accepted}

\abstract{We find a dimming during optical quiescence of the cataclysmic variable WZ\,Sge by about half a magnitude
between superoutbursts. We connect the dimming with the development of a cavity in the inner part of the
accretion disk. We suggest that, when the cavity is big enough, accretion of material is governed by the magnetic field of the white dwarf 
and pulsations from the weakly magnetic white dwarf appear.
The time scale of forming the cavity is about a decade, and it persists throughout the whole quiescent phase. 
Such a cavity can be accommodated well by the proposed magnetic propeller model for WZ\,Sge, where during quiescence mass is being expelled 
by the magnetic white dwarf from the inner regions of the accretion disk to larger radii.
}


   \maketitle
%

\section{Introduction}
\label{introduction}

Cataclysmic variables (CVs; for a review see Warner 1995) are binary systems wherein a white dwarf gains material from a gravitationally bound
companion star. Some of these CVs (called dwarf novae) show recurrent enhancements in brightness by several magnitudes (i.e., outbursts)
which come from enhanced accretion of material, onto the white dwarf through an accretion disk.
If the inclination of the binary is high enough, periodic increases in flux (`humps')
are discerned in the light curves when the system is at rest (i.e., in quiescence).
These humps have their origin in the region where the stream of material from the companion star interacts with
the accretion disk (so-called bright spot), and they appear when that site comes into our view, once per orbital period.

WZ\,Sge is an intriguing CV. It is an extreme member of a subclass
of the dwarf novae, called SU\,UMa stars. SU\,UMa stars show two kinds of outbursts:
1) normal outbursts, which typically last several days, and 2) superoutbursts with maximum brightness being about 
a magnitude brighter than normal outbursts, which last a few weeks (hence the prefix `super').
Superoutbursts occur less frequently than the normal outbursts.
During the main part of the superoutburst, periodic humps are discerned in the light curves
(irrespective of the binary's inclination) with a period that is a few percent longer than the orbital period; 
they are referred to as superhumps. The extreme members of the SU\,UMa stars usually do not show normal
outbursts, but only superoutbursts. These superoutbursts recur after very long intervals (years to decades) and show 
large amplitudes, typically five magnitudes or more, from quiescence to peak of the outburst. 
These systems are given yet another label: WZ\,Sge
stars (e.g., Bailey 1979) or TOADs (Howell et al.\ 1995).  

WZ\,Sge has an orbital period of 81.6\,min (Krzemi\'nski 1962, Warner 1976).
One sees it at an inclination of about 75$\degr$, just high enough for the donor star
to obscure the accretion disk but not the white dwarf (Krzemi\'nski 1962, Smak 1993).
So far it has displayed (super)outbursts in 1913, 1946, 1978 and 2001
(e.g., Mayall 1946, Patterson et al.\ 1981, 2002, Kato et al.\ 2009; see, e.g., Kuulkers et al.\ 2002, for the 
outburst light curves).

The standard disk-instability model (Osaki 1974), with some modifications (see, e.g., Lasota 2001,
Schreiber \&\ Lasota 2007 for discussions), can explain the normal outbursts of dwarf novae.
Two main competing models have been put forward for the superoutbursts.
On the one hand, superoutbursts are thought to occur
when the outer disk reaches the tidal radius, and enhanced tidal dissipation occurs
due to a resonance between the orbit of the outer disk and the orbit of the secondary star, 
increasing the mass transfer rate even more
(e.g., Osaki 1989, 1996; thermal tidal instability model). On the other hand, the enhanced mass transfer model predicts
a pure enhanced mass transfer from the secondary star,
owing to, e.g., irradiation of the secondary (see Hameury et al.\ 2000).
Also, a `hybrid' solution between the two has been suggested (e.g., Smak 2000).
For a review of the different flavours of outburst models we refer to Lasota (2001), among others.

Neither the thermal tidal instability model nor the enhanced mass transfer model
can fully explain what is happening in WZ\,Sge (see, e.g., Schreiber et al.\ 2004, Matthews et al.\ 2007,
for recent discussions). The observed recurrence times are quite long, and all outbursts 
are seen to start immediately as superoutbursts (similar to what is seen in low-mass X-ray 
binary transients, see, e.g., Lasota 2001). Somehow the inner disk regions need to be cleared 
during quiescence, in order to explain the observations
(Angelini \&\ Verbunt 1989, Lasota et al.\ 1995, Warner et al.\ 1996, Hameury et al.\ 1997, Meyer-Hofmeister et al.\ 1998,
Matthews et al.\ 2007).

Monitoring the quiescence is thus important for constraining modelling of the (super)outburst
behaviour. This can be a problem because most of the systems in quiescence are too faint to be observed by 
moderate equipment. However, since WZ\,Sge is nearby 
($\simeq$43\,pc; Thorstensen 2003, Harrison et al.\ 2004), it can rather easily be
observed during quiescence (V$\sim$15~mag, see also Sect.~\ref{results}). 
A huge amateur database exists on this source. Part of the available data has already been 
presented at the conference 
`The Physics of Cataclysmic Variables and Related Objects', in 2001, in G\"ottingen, Germany.
One of us (EK) showed the quiescence data and pointed out a decrease in brightness 
several years before the 2001 superoutburst. It was questioned, however,
whether this apparent drop 
may have been caused by a change of data acquisition methods of the amateurs
in the mid-90's (see Marsh 2002).
In this paper we come back to this issue and present data also beyond the 2001
superoutburst, as well as prior to the 1978 superoutburst. 

\section{Observations}
\label{data_analysis}

\subsection{Amateur observations}

The American Association of Variable Star Observers (AAVSO) holds a huge international
database of observations of variable stars by thousands of observers from all over the world,
dating back to its foundation over 90 years ago (e.g., Henden 2006). 
Visual estimates (by eye) of these stars typically have an uncertainty of about 0.2--0.3 magnitudes,
although some of the more experienced amateur observers can reach a precision down to about
0.02 magnitudes. The uncertainty in CCD observations varies typically between 0.1 to 0.01 magnitudes, 
but some professional and experienced amateur observers can obtain milli-magnitude precision
(see, e.g., Price et al.\ 2009). Observers are provided with a common
observing chart to ensure a homogeneous set of comparison and check stars 
(see also Appendix A). 
Additional information associated with individual observations may be available, including photometric
properties such as uncertainty, comparison stars used, and reports on seeing conditions.

All data go through strict quality control analysis (`validation'), a procedure that insures 
the data are as error free as possible. For WZ\,Sge, we selected all
visual and V-band observations that passed the validation tests. 
We did not use uncertain measurements or data which had
only undergone pre-validation; this resulted in 18751 measurements (excluding upper limits) up to 
2007 September 24, on the date of data extraction (2010 September 8). 233 observers contributed
to the dataset.

\subsection{RoboScope observations}

RoboScope is an autonomous 0.41-m telescope in central Indiana, USA, that is
dedicated to long-term monitoring of stellar accretion systems
(see Honeycutt et al.\ 1994, and references therein). 
Although the telescope was equipped with standard UBVRI filters, most of the time V-band only was used.
The working magnitude range of the system was about 11--18.
Generally, one or two 4 minute exposures per clear night were obtained for each of the programme 
stars, giving a data spacing of typically 1--4\,days.
The data are reduced using the method of incomplete ensemble photometry (Honeycutt 1992), which
uses all the stars as comparisons, even if the comparison stars do not appear on all 
the CCD exposures in the ensemble. 
The incomplete ensemble solution used 141 comparison stars, and the 
error on each WZ\,Sge's measured differential magnitude is about 0.03\,mag.
The resulting light curve has an arbitrary zero point, 
which is established to within 0.01\,mag using 14 secondary standard stars in the field
from Henden \&\ Honeycutt (1997).  
A transformation coefficient for the V filter (evaluated from regular observations of standard fields)
of +0.05 is applied in the zero-point calculation, assuming a B$-$V colour value of 0.0 (see Appendix A) for each
observation of WZ\,Sge.
There were 1078 usable RoboScope images of the field of WZ\,Sge, acquired during the interval November 1990 to 
December 2004 over 635 nights. 

\subsection{Multicolour Stiening photometry}
\label{Stiening}

High-speed multicolour photometry of WZ\,Sge was obtained with the Stiening photometer 
(see, e.g., Horne \&\ Stiening 1985) in 1982, 1988, 1991 (Skidmore et al.\ 1997),
and 1997 (Skidmore 1998). As far as we are aware, this dataset provides the only 
more or less homogeneous multicolour coverage over a single quiescent interval.
The Stiening photometer is a multi-channel photometer that uses 
dichroic mirrors to split the light beam into different channels. The different colours are 
measured simultaneously with separate photomultiplier tubes on each channel.
In 1982 the photometer provided three channels (UBR); later observations
provided UBVR photometry (plus an offset photometer for observing a nearby bright
star in white light to check for potential cloud and tracking problems). 
The UBR filters used in 1982 are not exactly the same as those used in the other years 
(see Skidmore et al.\ 1997, Skidmore 1998). Also, a new V tube was installed
in the photometer in January 1991, so the response differs slightly from previous observations
(see Wood et al.\ 1993). The Stiening UBVR passbands are not the same as the Johnson UBVR passbands, and
we, therefore, cannot directly compare these observations with the rest of our sample. However,
since we are only interested in relative light changes over the years, we decided not to attempt to convert the observed fluxes
into Johnson magnitudes. 

\begin{table} 
\caption{Log of observations of WZ\,Sge using the Stiening Photometer$^1$.}
\begin{tabular}{lc@{}c@{}c@{}cc}
\hline
\multicolumn{1}{c}{Date} &
\multicolumn{1}{c}{Band} & 
\multicolumn{1}{c}{Telescope} & 
\multicolumn{1}{c}{Ap.} & 
\multicolumn{1}{c}{$t_{\rm int}$} &
\multicolumn{1}{c}{$t_{\rm dur}$} \\
\hline
1982 Jun 23 & UBR  & 1.5-m Mt.~Lemmon &  17$\arcsec$ & 1.25\,s & 2.8\,hrs\\
1988 Aug 18 & UBVR & 2-m McDonald    &  6.7$\arcsec$ & 1.00\,s & 2.0\,hrs\\
1988 Aug 19 & UBVR & 2-m McDonald    &  6.7$\arcsec$ & 1.00\,s & 4.0\,hrs\\
1991 Aug 16 & UBVR & 2-m McDonald    &  6.7$\arcsec$ & 1.00\,s & 1.3\,hrs\\
1991 Aug 17 & UBVR & 2-m McDonald    &  6.7$\arcsec$ & 1.00\,s & 2.6\,hrs\\
1997 Aug 29 & UBVR & 2-m McDonald    & 26.8$\arcsec$ & 1.00\,s & 4.1\,hrs\\
1997 Aug 30 & UBVR & 2-m McDonald    & 26.8$\arcsec$ & 1.00\,s & 5.5\,hrs\\
1997 Aug 31 & UBVR & 2-m McDonald    & 26.8$\arcsec$ & 1.00\,s & 3.9\,hrs\\
1997 Sep 1  & UBVR & 2-m McDonald    & 26.8$\arcsec$ & 1.00\,s & 3.8\,hrs\\
\hline
\end{tabular}
\note{Ap. = aperture size used, $t_{\rm int}$ = integration time per measurement, $t_{\rm dur}$ = duration of
the observing run.
}
\label{log}
\end{table}

\begin{table*}
\caption{Nightly averaged Stiening pass-band fluxes of WZ\,Sge, including the errors in the average and an indication of the variability in terms of rms.}
\begin{tabular}{lccccccccc}
\hline
\multicolumn{1}{c}{Date} &
\multicolumn{1}{c}{JD} &
\multicolumn{1}{c}{U (mJy)} &
\multicolumn{1}{c}{rms} & 
\multicolumn{1}{c}{B (mJy)} & 
\multicolumn{1}{c}{rms} & 
\multicolumn{1}{c}{V (mJy)} & 
\multicolumn{1}{c}{rms} & 
\multicolumn{1}{c}{R (mJy)} &
\multicolumn{1}{c}{rms} \\
\hline
1982 Jun 23 & 2445144 & 3.482$\pm$0.003 & 0.28 & 3.569$\pm$0.003 & 0.25 & -- & -- & 4.007$\pm$0.005 & 0.46 \\
1988 Aug 18 & 2447392 & 3.383$\pm$0.003 & 0.25 & 2.981$\pm$0.002 & 0.17 & 2.883$\pm$0.002 & 0.20 & 2.567$\pm$0.002 & 0.20 \\
1988 Aug 19 & 2447393 & 3.475$\pm$0.002 & 0.30 & 3.055$\pm$0.002 & 0.21 & 2.676$\pm$0.002 & 0.21 & 2.480$\pm$0.002 & 0.20\\
1991 Aug 16 & 2448485 & 3.142$\pm$0.003 & 0.21 & 3.198$\pm$0.002 & 0.15 & 3.088$\pm$0.002 & 0.16 & 2.799$\pm$0.003 & 0.20\\
1991 Aug 17 & 2448486 & 2.688$\pm$0.002 & 0.20 & 3.031$\pm$0.002 & 0.16 & 3.000$\pm$0.002 & 0.18 & 2.775$\pm$0.002 & 0.22\\
1997 Aug 29 & 2450690 & 2.785$\pm$0.004 & 0.20 & 2.829$\pm$0.003 & 0.16 & 3.088$\pm$0.004 & 0.21 & 4.104$\pm$0.010 & 0.47\\
1997 Aug 30 & 2450691 & 2.942$\pm$0.002 & 0.23 & 2.933$\pm$0.002 & 0.27 & 3.085$\pm$0.004 & 0.60 & 2.073$\pm$0.007 & 0.99\\
1997 Aug 31 & 2450692 & 2.618$\pm$0.005 & 0.21 & 2.653$\pm$0.004 & 0.15 & 3.168$\pm$0.006 & 0.26 & 3.333$\pm$0.014 & 0.57\\
1997 Sep 1  & 2450693 & 2.672$\pm$0.005 & 0.20 & 2.748$\pm$0.004 & 0.14 & 3.548$\pm$0.007 & 0.28 & 2.840$\pm$0.011 & 0.42\\
\hline
\end{tabular}
\label{table_obslog}
\end{table*}

The multicolour Stiening observation log is given in Table~\ref{log}.
Measurements from the photometer were continuously recorded and the telescope was moved between target,
sky, and comparison stars. All the data were taken during dark time. 
Sky measurements were made before and after the observing run in 1982;
they were done during the observing runs approximately every 10 to 15 minutes in 1988 and 1991, 
and between 15 to 40 minutes in 1997. The target and comparison star data were sky-subtracted
in each passband after interpolating between sky observations using a cubic spline fit.
The data were then corrected for extinction. Flux calibration was carried out using observations of 
standard stars. A red field star at about 10$\arcsec$ from WZ\,Sge\footnote{The angular distance is given in the 
J2000.0 coordinate frame (see Harrison et al.\ 2009, 
where the red field star is referred to as `Ref-5'). Krzemi\'nski (1962) and Krzemi\'nski \&\ Kraft (1964) 
quote distances of 8$\arcsec$ and 7$\arcsec$, respectively. This is because the mean proper motion of
WZ\,Sge is about 0.08$\arcsec$ per year (van Maanen 1926, Thorstensen 2003, Harrison et al.\ 2004), whereas
the mean proper motion of the closeby red field star is consistent with 0$\arcsec$ (Harrison et al.\ 2004),
so, in about 40 years the angular distance between the two stars has indeed increased by about 3$\arcsec$.}
was included in the entrance aperture of the photometer in 1997, 
owing to poor seeing conditions. During the 1982, 1988, and 1991 observing runs, WZ\,Sge was
slightly offset in the aperture to avoid including this nearby star.
Separate measurements of this nearby star were made on August 31, 1997, and
the contaminating star flux was subtracted from the 1997 main target light curves. 
For a more detailed description of the data reduction we refer to
Skidmore et al.\ (1997) and Skidmore (1998). 
The nightly averaged Stiening passband fluxes are given in Table~\ref{table_obslog}, together with
the standard error in the mean (1$\sigma$). Additionally, we list the rms values to give an idea of the source variability
within a time bin. In such a time bin the source variability is mainly due to orbital variations.
The various observing runs sample between one to four whole orbital periods each (see Table~1), so any bias in the nightly averages due to unequal sampling 
of orbital phases is expected to be small. The colour variations within an observing run show a combination of slight orbital modulation 
and random fluctuations (see Skidmore 1998). Any bias there is typically less than about 0.1\,mag, which is very near the systematic 
orbital colour modulations.

\subsection{Additional measurements from the literature}

Because of the sparse sampling of the AAVSO data before the 1978 superoutburst 
we augmented our sample with visual and photometric V-band observations reported in the literature between 1946 and 
1978 (Lohmann \&\ Miczaika 1946, Ahnert 1947, Steavenson 1947, 1948, 1950, 1953, 
Beyer 1952, Walker 1957, Krzemi\'nski \&\ Smak 1971). We did not use their upper limits or any uncertain measurements.
For completeness we also included photometric V-band observations taken during the 1978 
(Targan 1978a,b, 1979, Heiser \&\ Henry 1979, Bruch 1980, Howarth 1981, Patterson et al.\ 1981)
and 2001
(Patterson et al.\ 2002, Howell et al.\ 2004)
superoutbursts.

Many photographic plates exist from before the 1913 superoutburst up to several years after the 1946 superoutburst. 
We used the reported 
(Mayall 1946, Whitney 1946, Lohmann \&\ Miczaika 1946, Ahnert 1947, Bigay 1947, 1948, Rosino 1948, 1952; see also Eskio\v glu 1963) 
photographic magnitudes to review WZ\,Sge's behaviour around and in between the 1913 and 1946 superoutbursts.
We note that these early photographic plates were somewhat more sensitive to blue than to red light.

\section{Results}
\label{results}

\begin{figure*}[top]
\centering
  \includegraphics[height=.55\textheight,angle=-90]{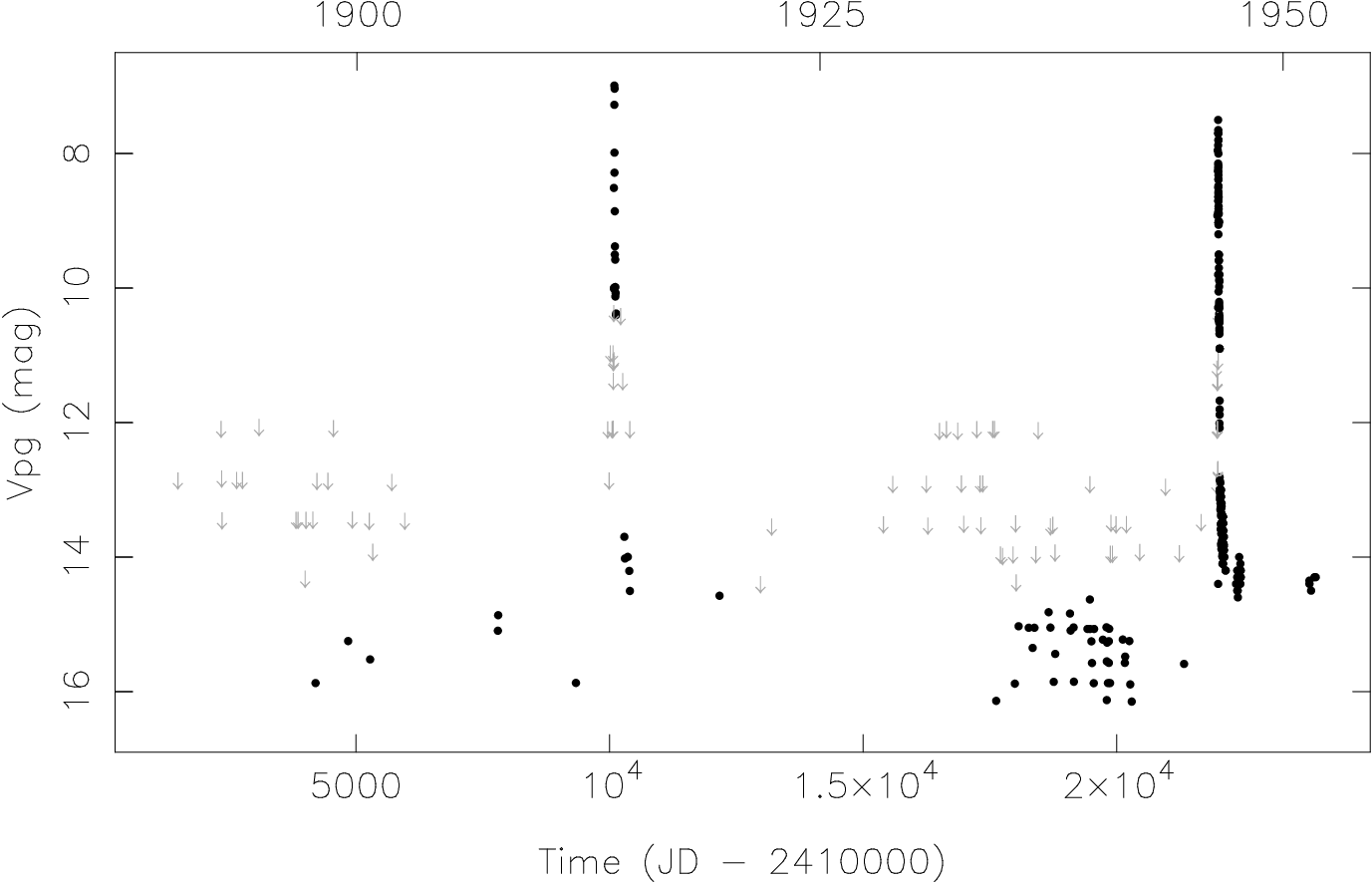}
~\\
  \includegraphics[height=.57\textheight,angle=-90]{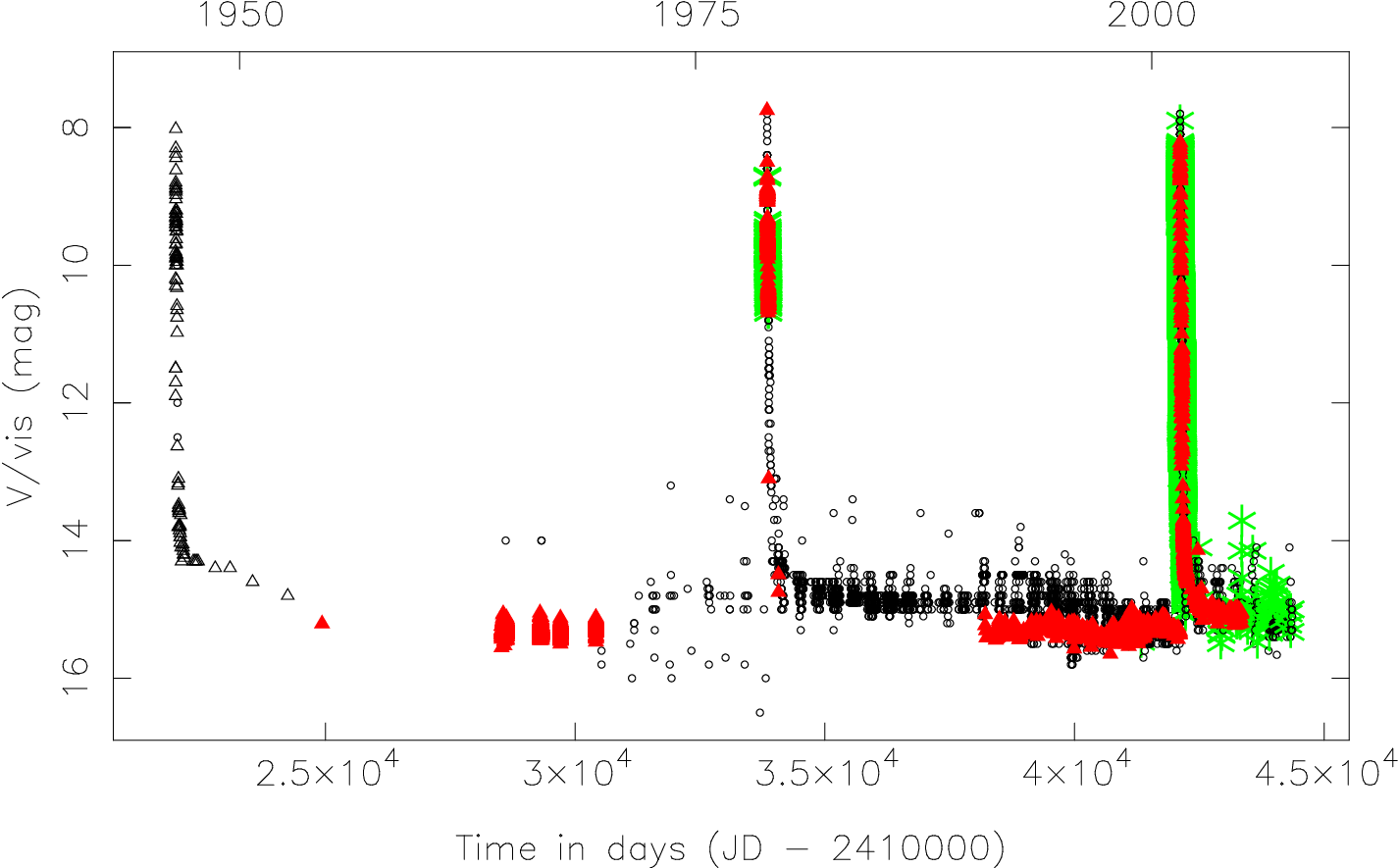}
  \caption{
{\it Top}: All photographic plate measurements between 1890 and 1951 available in the literature.
Upper limits are indicated with light grey arrows.
{\it Bottom:} All visual (small open circles) and photometric V-band (green asterisks) detections available in the AAVSO database
that passed the AAVSO validation tests. Also plotted are visual (open triangles) and V-band observations
available in the literature, as well as the RoboScope V-band data (all red filled triangles).
}
\label{all_aavso_roboscope_wzsge}
\end{figure*}

The photographic measurements are displayed 
in the top panel of Fig.~\ref{all_aavso_roboscope_wzsge}.
All other observations (except the Stiening photometry) 
are shown in the bottom panel of Fig.~\ref{all_aavso_roboscope_wzsge}.
The superoutbursts that occurred in 1913, 1946, 1978, and 2001 are clearly visible as strong increases by up to 7 magnitudes.
Sparse observations in quiescence exist before the 1913, 1946, and 1978 superoutbursts. 
The quiescent interval between the 1978 and 2001 superoutbursts is well
covered, as is the period after the 2001 superoutburst.
Generally, in quiescence, WZ\,Sge's optical magnitude is roughly between 14 and 16, with an average of 
about 15 mag. The spread in the visual (and probably the photographic) data points by more than about 0.5\,mag is mainly caused by the different 
conditions in which the visual observations were done, as well as, for example, differences in the eye's 
sensitivity and telescope system used between the various observers (e.g., Zissell 2003; see Appendix A). 
This is larger than the typical orbital variations, which are caused by the orbital hump and partial disk eclipse,
as is shown by, e.g., the 1964--1969 dataset from 
Krzemi\'nski \&\ Smak 1971 (see Fig.~\ref{all_aavso_roboscope_wzsge}). 

During a superoutburst WZ\,Sge fades on a time scale of weeks to a month
down to an optical magnitude of around 14.5
(see, e.g., Patterson et al.\ 1981, Kuulkers 2000, Kuulkers et al.\ 2002, for a comparison
of various superoutburst light curves).
About a year after 
the 1913 and 1946 superoutbursts, the average photographic magnitudes are 14.1$\pm$0.1 and 14.30$\pm$0.03, respectively.
The latter value is consistent with the visual magnitude reported by Steavenson (1948) around that time.
We find yearly visual averages of 14.70$\pm$0.02 and 14.83$\pm$0.03, just after the 1978 and 2001 superoutbursts, 
respectively. The V-band data are consistent with these values. A few years after the 1913, 1946, and 1978 superoutbursts
WZ\,Sge is still at a similar flux level to the one quoted above.

During the decade before the 1913 (from about 1897 to 1912) and 1946 (from about 1934 to 1945) superoutbursts the 
data available provide average photographic magnitudes of 15.4$\pm$0.2 and 15.4$\pm$0.1, respectively.
About eight years after the 1946 superoutburst, it was observed at V$\simeq$15.2\,mag (Walker 1957). The available
data indicate a gradual fading over the years after the 1946 superoutburst. During the 1964--1969
time period (Krzemi\'nski \&\ Smak 1971), the average V magnitude was 15.279$\pm$0.002, 
suggesting the source had not declined in brightness. The visual observations between 1969 and 1978 
show an average of 15.0$\pm$0.1, so again there is no further indication of a change in the optical flux. 

\begin{figure}[top]
\centering
  \includegraphics[height=.34\textheight,angle=-90]{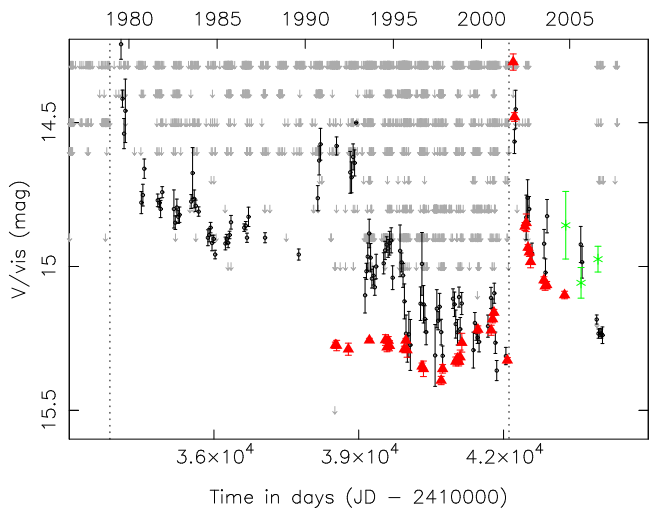}
  \caption{
Monthly averages of all the visual (small open circles) and RoboScope V-band (red filled triangles) 
data displayed in Fig.~\ref{all_aavso_roboscope_wzsge} (bottom),
as well as yearly averages (because of the few measurements available) for the AAVSO V-band data (green asterisks).
Shown are the 1978--2001 quiescent and post-2001 superoutburst intervals.
Only those points whose averages are based on more than 11 (arbitrary chosen number) measurements are shown. 
Errors shown are the 1$\sigma$ uncertainties per interval. 
Upper limits are given by light-grey arrows. (Many upper limits after 2003 have not been 
validated yet, which explains the apparent gap in that time frame).
The times of the peak of the 1978 (near JD\,2443845) and 2001 (near JD\,2452115) superoutbursts are marked by dotted lines.
}
\label{zoom}
\end{figure}

After the 1978 superoutburst, WZ\,Sge declines in brightness slowly up to 1990 to a yearly average magnitude 
of 14.92$\pm$0.02 (see also Fig.~\ref{zoom}). From 1990 up to the end of our dataset, we have contemporaneous and more or less continuous coverage
of visual and V-band observations. As can be seen from the bottom panel of Fig.~\ref{all_aavso_roboscope_wzsge}, the 
visual measurements are on average lower in magnitude than the RoboScope V-band measurements, in the 
1990--1997 time frame. From about 1997 onwards the visual and V-band measurements (RoboScope, AAVSO and other) agree, however.
To investigate the apparent discrepancy in the 1990-1997 time frame, 
we determined monthly averages of our dataset, where we included only time bins with more than 11 (arbitrary) estimates. 
These monthly averages are shown in Fig.~\ref{zoom}.
The discrepancy varies between about 0.15 (1996--1998) and 0.65 magnitudes (1990--1993).
As discussed in more depth in Appendix~A, we cannot say with certainty what exactly causes the large offset.
However, because the RoboScope V-band data have all been analysed in a homogeneous way, using 
many different comparison stars, and because various visual upper limits show a lower brightness than is
indicated by the visual detections in the time frame where the discrepancy is seen, we attribute the difference to mainly 
(possibly cumulative) systematic effects in
the AAVSO data (see Appendix~\ref{systematics}).\footnote{The apparent drop in brightness several years before the superoutburst,
as presented at the CV conference in 2001 (see Sect.~1), may thus indeed be regarded as unrelated to WZ\,Sge itself.}
Any conclusions about the source behaviour in this period is, therefore, based on the RoboScope data. 

The RoboScope V-band data (Figs.~1, bottom and 2) show 
a low-amplitude (few tenth of a magnitude) 
modulation with a time scale of about three years. 
The average of the RoboScope data in that time frame is V=15.300$\pm$0.003,
i.e., similar to what was observed by Krzemi\'nski \&\ Smak (1971).
We note that a low-amplitude modulation with a similar period seems also to be present during the decline
in visual flux from 1979 to 1990 (see Fig.~\ref{zoom}). Just before the 2001 superoutburst and at the end of our dataset
(September 2007), the yearly average visual magnitude was 15.17$\pm$0.02 and 15.22$\pm$0.02, respectively.

The only multi-band information available over a large part of the 
1978--2001 quiescent interval comes from the (sparse) Stiening photometry 
(Skidmore et al.\ 1997, Skidmore 1998; see Table 2). 
In the 1988--1997 period there is no evidence that the source dims
(consistent with the RoboScope dataset), except maybe for the 
Stiening U-band\footnote{A decrease in the ultraviolet light was also reported by Krzemi\'nski \&\ Kraft (1964), between
about JD\,2437550 (September 1961) and JD\,2437940 (October 1962) by about 0.2 magnitudes, almost two decades before the 1978 superoutburst.}. 
In the Stiening V and B-band the source's brightness may even be increasing
on a long-term time scale, but due to the sparse time coverage we cannot rule out that
this is caused by a similar low-amplitude modulation as seen in the RoboScope data.
In the Stiening B and R-bands, the source seemed to have declined in brightness between 1982 and 1988.

We conclude that for several years after a superoutburst the average optical brightness of WZ\,Sge is higher by about
half a magnitude with respect to what was measured later on in quiescence. The available data suggest that it takes about a decade
or so to reach the final quiescent level after at least the 1946 and 1978 superoutbursts. After the 2001 superoutburst, it took
only about half a decade to reach the same final quiescent level.

\section{Discussion}
\label{discussion}

\subsection{The dimming and the development of a hole in the disk}

We find evidence for a dimming in WZ\,Sge by about 0.5\,mag during the quiescent interval between superoutbursts.
The dimming seems to take place on a time scale of about a decade. For the rest of the 
quiescent interval, the brightness stays more or less constant at an average of V$\simeq$15.3\,mag.
There is some evidence that the dimming period after the 2001 superoutburst is shorter by a few years.
We speculate that the latter may have to do with the quiescent period before
the 2001 superoutburst lasting about 10 years shorter than the quiescent periods before the 1946 and 1978 superoutbursts.

The standard, as well as modified, versions of the disk-instability model predict a brightness 
{\it increase} by typically 0.5--1\,mag during the quiescent interval in between normal outbursts 
(see, e.g., Meyer \&\ Meyer-Hofmeister 1984, Mineshige 1986, Duschl \&\ Livio 1989; see also Smak 2000). 
Assuming this model would account for superoutbursts as well, it
contradicts our observations and, therefore, WZ\,Sge's superoutbursts 
cannot be triggered by a `pure' disk-instability. 
Smak (2009, and references therein) have recently discussed that superoutbursts are due to a 
major enhancement in the mass-transfer rate.
The standard mass-transfer instability model (e.g., Bath 1975) would give 
a steady or slight decrease in brightness during the quiescent interval in between 
(normal) outbursts, since the brightness reflects the mass transfer rate from the secondary.
That we do not see evidence of any increase in brightness over the course of quiescence,
therefore, seems to favour an enhanced mass transfer model. However, we note that recent model
calculations of various versions of the enhanced mass transfer model also predict
brightness increases between outbursts (see, e.g., Schreiber et al.\ 2004).

Since the accretion disk is not the sole contributor to the optical 
light\footnote{Krzemi\'nski \&\ Smak (1971) estimated the relative quiescent contributions from the white dwarf,
bright spot, and accretion disk in the V-band as 70--80\%, 15\%, and 5--15\%, respectively.}, a way out
may be that the other components (white dwarf and bright spot) account for the 
total light to decrease over the course of the quiescent interval. 
However, observations of, for example, the eclipsing dwarf nova Z\,Cha have shown that 
the white dwarf fades within weeks after the (normal) outburst and then stays 
at the constant level, while the bright spot luminosity first increases and then
also stays constant. The accretion disk flux did not show evidence for
a secular trend in between outbursts (van Amerongen et al.\ 1990).
In WZ\,Sge, the white dwarf cools the most during the first year after the superoutburst;
later, the cooling is much slower. Also, most of the white dwarf's
contribution is in the UV, and as it cools, the peak of the brightness distribution
shifts towards optical wavelengths (la Dous 1994; Godon et al.\ 2006). In the optical, therefore, 
the (initial) amount of dimming of the white dwarf is expected to be less. We thus conclude that the 
white dwarf is not the cause of the overall dimming in the quiescence of WZ\,Sge. Similarly, the bright
spot is also not expected to cause any overall dimming (but rather a brightening), if it shows 
the same trend as inferred for Z\,Cha. We, therefore, identify the disk as the origin of the
fading seen in WZ\,Sge.

The dimming we see corresponds to a change in flux by about a factor of 1.5. 
`Holes' in the disk have been invoked to explain the long recurrence
times between the superoutbursts in WZ\,Sge stars (see Sect.~1).
We suggest that the decrease in brightness is 
due to the removal of the inner disk regions during the quiescent phase. This removal 
may be done by the magnetic propeller operating effectively in WZ\,Sge (Patterson et al.\ 1998, Lasota et al.\ 1999, 
Matthews et al.\ 2007; see also Sect.~4.2).
If this is the case, and assuming the optical emission at the end of a superoutburst is coming
from a full circular flat disk, while the one before a superoutburst comes from a circular
flat ring, one can infer that (and assuming the average disk temperatures and the relative contributions
to the overall light are the same):
$L_{\rm circ}/L_{\rm ring} = R^2_{\rm out,1}/(R^2_{\rm out,2}-R^2_{\rm in,2})$, where
$L_{\rm circ}$ and $L_{\rm ring}$ are the luminosity of a circular disk and a ring, respectively,
$R_{\rm out,1}$ and $R_{\rm out,2}$ the outer disk radius at the end and before a superoutburst,
and $R_{\rm in,2}$ the inner disk radius just before a superoutburst\footnote{For a full circular disk we 
take for simplicity that the inner disk radius at the end of a superoutburst, $R_{\rm in,1}$, is much smaller
than $R_{\rm out,1}$. In practice, $R_{\rm in,1}$$\sim$$R_{\rm wd}$, where $R_{\rm wd}$ is the radius of the 
white dwarf, i.e., $R_{\rm in,1}$ is more than a magnitude smaller than $R_{\rm out,1}$ (see, e.g., Lasota et al.\ 1999).}.
If we assume the outer disk radius to be constant (e.g., Mason et al.\ 2000; but see below), 
i.e., $R_{\rm out,1} = R_{\rm out,2} = R_{\rm out}$,
one can then show that before a superoutburst $R_{\rm in}/R_{\rm out}$ is about 0.6.
This is close to what is estimated by Lasota et al.\ (1999) for WZ\,Sge in quiescence, i.e., $\simeq$0.7.
If, however, the outer disk radius increases during quiescence (see, e.g., Lasota 2001; but see below), 
it lowers the value of $R_{\rm in}/R_{\rm out}$. Also, the average temperature in a circular disk
is expected to be higher than for a ring-like disk, and this also has the effect of lowering
the value of $R_{\rm in}/R_{\rm out}$. These two effects may bring the value closer to
what is measured from optical emission-line spectra, i.e., about 0.2--0.3 (Mennickent \&\ Arenas 1998, Mason et al.\ 2000).
WZ\,Sge's high inclination, changing the viewing geometry of a full and ring-like disk, may also affect our simple estimate.

Changes in the accretion disk geometry are expected to affect the shape of emission lines in the 
spectra. Spectroscopic observations have already indicated changes in the hydrogen emission-line profiles
a year and a half after the 1978 superoutburst with respect to five months before the
1978 superoutburst (Voikhanskaya 1983). A first attempt at a systematic comparison of optical spectra throughout
WZ\,Sge's quiescence was made by Neustrov (1998), by deriving outer disk radial information from
the 1946--1978 and 1978--2001 quiescent phases. It was concluded that $R_{\rm out}$
{\it decreases} by about 15\%\ during the quiescent phase. From a similar 
in-depth study, Mason et al.\ (2000) conclude, however, 
that $R_{\rm out}$ does {\it not} vary significantly during quiescence.

The expectations from the thermal tidal instability model are different (see, e.g., Lasota 2001).
In between superoutbursts, the thermal tidal instability model requires the disk to grow (i.e., 
an increase in $R_{\rm out}$) after every normal outburst, until it reaches the
diameter at which there is a 3:1 resonance between the period of the 
outer disk and the secondary star. No normal outbursts are seen in WZ\,Sge, which
is another argument against the thermal tidal instability model for this system. Owing to the 
redistribution of angular momentum during quiescence of WZ\,Sge, however, the disk
is still expected to grow (Meyer-Hofmeister et al.\ 1998).
The conclusions by Neustrov (1998) and Mason et al.\ (2000), as well as disk radius 
measurements in other eclipsing dwarf novae in between (normal) outbursts (i.e., support for a 
shrinkage of the outer disk in, e.g., U\,Gem: Smak 1971, 2001; Z\,Cha: O'Donoghue 1986),
thus contradict the above expectations. This may be further support for the enhanced mass transfer model.
Direct evidence of enhanced mass-transfer during WZ\,Sge's 2001 superoutburst was
presented by Patterson et al.\ (2002) and Steeghs (2004).

\subsection{WZ\,Sge as a magnetic propeller}

WZ\,Sge is considered to be weakly magnetic (Patterson et al.\ 1998, Lasota et al.\ 1999, and 
references therein). If there is a hole in quiescence,
accretion onto the white dwarf may occur predominantly along the magnetic field lines, and pulsations are expected.
Interestingly, in the two years preceding the 1978 superoutburst, a stable 
27.87\,s pulse period was found in the optical 
(Robinson et al.\ 1978, Patterson 1980). After the 1978 outburst it did not reappear, until 
only in 1995 it was found again, also in X-rays 
(Patterson et al.\ 1998, Skidmore et al.\ 1999, and references therein). Similarly, no evidence for
27.87\,s pulsations were found in the optical and X-rays, almost two years
after the 2001 superoutburst (Mukai \&\ Patterson 2004).
It has been suggested that the 27.87\,s pulse period is caused by the rotation of the white dwarf. However, the 
difficulty in finding it at various times was a problem for the interpretation
(see the discussion in Patterson et al.\ 1998). 
We suggest connecting the appearance of a significant pulse signal years before a superoutburst with
the observed dimming of the system. As a cavity is created in the inner disk regions around the white dwarf,
the region clears up, accretion starts to become dominated by the magnetic field, and subsequently, pulsations
become observable.
We note that there are various other (transient) (quasi-periodic) oscillations, with periods close to the 27.87\,s pulses
(see, e.g., Patterson et al.\ 1998, Skidmore et al.\ 1999, Lasota et al.\ 1999).
They can be consistently accommodated if the white dwarf is magnetically important enough, 
within the magnetic propeller model (e.g., Lasota et al.\ 1999, Matthews et al.\ 2007,
Warner \&\ Pretorius 2008).

Only a few models that invoke holes in the accretion disk make predictions
regarding the time scale for the existence of the holes. 
The hole is expected to be formed just after the end of an outburst, when
most of the inner disk material has been accreted onto the white dwarf.
In WZ\,Sge, the hole is thought to be closed within a few years.
Evaporation at the inner disk may further lengthen the existence of the hole, but it is 
expected that the inner disk radius reaches the white dwarf within about a decade
(Meyer-Hofmeister et al.\ 1998). We, however, conclude the opposite; i.e., 
the hole grows in between superoutbursts. We assume that this takes place on the 
long time scale between superoutbursts (at least of the order of a decade). 
The existence of such a hole, far into quiescence, especially
when the spinning white dwarf is also seen, is supported by the 
various optical spectroscopic observations of WZ\,Sge mentioned above (Mennickent \&\ Arenas 1998, Mason et al.\ 2000).

Models that explain the long superoutburst recurrence time of WZ\,Sge by the action of a magnetic propeller
within the disk (e.g., Lasota et al.\ 1999, Matthews et al.\ 2007) predict the 
growth of a hole in quiescence. During a superoutburst, the magnetic propeller is 
quenched and the disk extends down to the white dwarf surface. At the end of the superoutburst, the 
disk returns to its `cold' state and the viscosity drops (within the framework of the disk-instability model), 
enabling the magnetic propeller to clear the 
inner regions of the disk by expelling mass to larger radii.\footnote{As mentioned by Lasota et al.\ (1999), 
not all material is expelled, though. Some of it will manage to funnel onto the white dwarf and produce 
the spin period pulsations.} 
As a result, a hole grows in the central regions of the disk, and it does {\it not} close up within a few years, as previously suggested.
Since the inner regions of a disk are hotter and mainly emit at shorter wavelengths than in the 
outer disk regions, the dimming should be more significant at shorter wavelengths (assuming
white dwarf cooling does not compensate at all). Also, the time scale for the clearing increases with disk radius 
so that the hottest regions of the disk should clear most quickly. Furthermore, emission at long wavelengths
may even increase as mass is added to the outer disk where dissipation increases.
This is indeed suggested by the multicolour observations presented in this paper. Further support comes
from X-ray observations that show that $\sim$2 years after the 2001 superoutburst WZ\,Sge was brighter 
by about a factor of 2.4 than it was $\sim$18 years after the 1978 superoutburst (Mukai \&\ Patterson 2004).

The material is thought to be propelled outwards to the radius at which the torque imparted on the 
disk by the magnetic propeller is balanced by the cold-state viscous torque (Matthews et al.\ 2007). 
This defines the radius of the hole within the disk. 
Possibly, a fraction of the expelled material may travel even farther and create the observed (Howell et al.\ 2008)
cool circumstellar dust ring.
The time scales associated with the magnetic and viscous 
torques increase with radius, so the time over which the hole forms within the disk can be
estimated by the viscous time at the outer radius of the hole:
\begin{equation} 
t_{\rm visc, cold} \sim \frac{R_{\rm hole}}{\alpha_{\rm c} c_{\rm s}} \frac{R_{\rm hole}}{H},
\end{equation}
where $\alpha_{\rm c}$ is the cold-state Shakura-Sunyaev (1973) viscosity parameter, $c_s$ is the sound speed, and $H$ the 
disk scale height. 
Associating this time scale with the observed time needed for the hole to develop (about 10 years), we can make an order of magnitude estimate
of the cold-state viscosity parameter. Assuming $R_{\rm hole}\sim 10^{10}$\,cm (e.g, Lasota et al.\ 1999, Matthews et al.\ 2007), 
$c_s\sim 5$\,km\,s$^{-1}$, and $H\,/R\sim 0.01$ (e.g., Matthews et al.\ 2007) leads to 
$\alpha_{\rm c}\sim 0.006$. This is very encouraging, as this is comparable to the canonical value of 
the cold-state viscosity adopted in models for the outbursts of dwarf novae.
Thus, the magnetic propeller 
model is able to explain the long recurrence time of WZ\,Sge, the existence of a hole in the central regions of the 
disk during quiescence, and the time scale on which the hole is formed, all while avoiding the imposition of an 
anomalously low cold-state viscosity required by other models (see, e.g., Smak 1993). 

To conclude, a more elaborate evaluation of the 
current existing models is necessary; e.g., clear brightness evolution predictions during the quiescent phase from the various
superoutburst models are needed, especially for the thermal tidal instability model
and the enhanced mass transfer model. It is not clear that they are similar
to what we quote above for what is expected in between normal outbursts.
Also, precise predictions of the evolution of the hole during quiescence are lacking.
Moreover, a detailed comparison of the quiescent properties (broad-band flux evolution, orbital light curve changes,
as well as disk radius variations during the quiescent intervals) ought to be done of the various observed WZ\,Sge stars. 
For example, (partial) eclipse observations in quiescence may help to highlight the evolution of the outer disk with respect to the inner regions and 
the white dwarf. This is, however, beyond the main scope of this paper.
We note here that, alternatively, further evidence for a hole in the disk may come from analysis 
of power spectra characterizing the fast (seconds) X-ray and optical variability. The breaks in the power 
spectra trace the truncation radius at the inner part of the accretion disk (Revnivtsev et al.\ 2010).

\section{Summary}

Using a data base of more than 60 years of optical observations of the dwarf nova WZ\,Sge, we sampled (part of) five long quiescent
phases between several superoutbursts. We find that just after a superoutburst the system is brighter in the optical
by about half a magnitude compared to what is seen just before a superoutburst. 
The time scale of decline in brightness during quiescence is about a decade. The dimming is consistent with
the development of a hole in the inner part of the accretion disk. We suggest that once the cavity has been cleared out,
the spin period of the white dwarf starts to be observable. Holes in the disk have been invoked in various (super)outburst
models to explain the long quiescent phase. However, most of these models predict a filling up of the hole
within a decade, whereas ours and previous measurements suggest the presence of such a hole throughout the whole quiescent phase.
The only model that predicts the development of a hole in quiescence is the magnetic propeller model, recently put forward for WZ\,Sge.
This is done by having matter being expelled to larger radii due to the relatively fast rotating, weakly magnetic, white dwarf.
We confirm that this model quantitatively describes basically all aspects of the quiescent phase well.
We, therefore, predict that soon, i.e., once a new cavity has been cleared out, pulsations from the spinning white dwarf
will appear again with the 27.87\,s period. 

\begin{acknowledgements}
We acknowledge with thanks the variable star observations from the AAVSO International 
Database contributed by observers worldwide and used in this research.
EK thanks Keith Horne and Rob Robinson for discussion of the Stiening photometer observations,
Joe Smak and Jean-Pierre Lasota for discussions of the disk changes, Mike Revnivtsev for pointing out an
alternative way to verify the presence of a hole in the accretion disk, and Doug Kniffen for discussions
of the Purkinje effect and the way an observer's eye works.
The use of information from Wikipedia, the free encyclopedia, about the eye's sensitivity is also acknowledged.
The comments by the referee were highly appreciated, since it led to a significant improvement in the current paper.
\end{acknowledgements}

\appendix

\section{Systematics in the AAVSO dataset?}
\label{systematics}

\subsection{Discrepancy between the AAVSO visual and RoboScope V-band observations}
\label{discrepancy}

Figure \ref{zoom} shows that the visual magnitude declines from about 14.5 to about 15.0 in the period 1979 to 1990,
with a possible low-amplitude modulation as present in the RoboScope dataset seen later (see Sect.~3). In 1990
the visual data show an apparent discontinuity, where the brightness increases by about 0.4\,mag. In the 1990--1993
time frame, the discrepancy between the visual and RoboScope V-band data is up to about 0.7\,mag. In 1993
the brightness decreases again sharply by about 0.4--0.5\,mag, with discrepancies between the visual and V-band
ranging from 0.15 to 0.35\,mag. In 1995 another sharp drop in visual magnitude can be discerned by about 0.3\,mag,
which made the visual measurements agree with the V-band data. However, during the two years after that, an offset of about 0.2\,mag
was again present. After 1997 the visual and V-band data generally agree.

The RoboScope data were analysed in a homogeneous way, using many different comparison stars. Moreover, wherever available, other
V-band data agree with the RoboScope measurements. 
Comparisons of AAVSO data and RoboScope V-band data for other variables show good agreement in general, also
in the 1990--1997 time frame (e.g., Honeycutt et al.\ 1998).
Also, typically, the visual upper limits shown in Fig.~\ref{zoom}
are consistent with the reported visual detections, except for the 1990--1993 period when there are various visual
upper limits pointing to the source being less bright than actually indicated by the detections.
We thus suspect systematics were present in the AAVSO data in the 1990--1997 period.
However, because of this, some doubts could be raised about the quality of our whole dataset, and, therefore, on our
findings in general.

It has long been recognized that visual estimates can differ from data taken in the (Johnson) V-band (see, e.g., Zissell 2003). 
Usually, it can be traced to differences in the comparison stars sequences used, reduction procedures, or other obvious culprits
(such as bandpass, sky background, or type of telescope system used).
Less obvious cases have sometimes been traced to probable colour effects, such as visual observers using multiple comparison 
stars of sometimes very different colours (from each other, as well as from the variable star) and V-band observers usually using only one comparison star.

In this appendix we, therefore, address the following questions. 
1) Are the apparent declines in brightness of WZ Sge in the years 
leading up to the 1978 and 2001 superoutbursts and, after the 2001 superoutburst, real,
or are they caused by systematic effects, such as changes in the 
observing sequence, protocol, individually contributing observers, or colour effects? 
2) Why is there a discrepancy between the AAVSO visual and RoboScope V-band data in the 1990--1997 period?

\subsection{Observing sequence and protocol}

The original AAVSO charts for WZ\,Sge were created by Wayne Lowder and Clinton Ford in October 1970, 
using the sequences in Lohmann \&\ Miczaika (1946), Krzemi\'nski \&\ Kraft (1964), 
and measurements by Charles Scovil using the Yale photometer. They contained one 14th-magnitude star 
northwest of WZ\,Sge, as well as a fainter, 15th-magnitude star. 
In 1973 a second revision was made, adding three 14th-magnitude stars from John Isles of the British 
Astronomical Association (BAA) very close to WZ\,Sge.  The addition of 
these stars allowed visual observers to make more accurate faint observations of 
WZ\,Sge, and enabled more observers to try to estimate the star's magnitude at minimum.
In December 1978, a third revision was made, using the Webbink photometry (see Bateson et al.\ 1981).
Some brighter comparison stars were changed, but none of the 14th- or 15th-magnitude stars near WZ\,Sge were changed. 
Thus, despite the Webbink photometry, the faint comparison stars did not change in the five years before and after 
the 1978 superoutburst. 
No changes were made to the visual observing protocol in all these years.

After the revision using the Webbink photometry, there were no other changes to the official AAVSO sequence 
before the 2001 superoutburst. The Royal Astronomical Society of New Zealand (RASNZ) created charts for WZ\,Sge 
in 1981, using the AAVSO chart and Webbink photometry. It should be noted that some of the RASNZ, as well as the
BAA, comparison stars' magnitude estimates differ by a tenth or two from the AAVSO magnitudes, even though they were all taken 
from the same reference.  
Again, there were no changes to the visual observing protocol. 
We, therefore, regard any long-term trends seen in WZ\,Sge in the 1970--2001 time frame to be 
unrelated to any change in observing sequence and protocol.

In 1997 Henden \&\ Honeycutt published secondary photometric standards for northern CVs and related objects, including
WZ\,Sge. The AAVSO sequence for WZ\,Sge, especially at the faint end, was changed during the 2001 superoutburst, since it was updated with 
the information provided by them. Numerous 14th- and 15th-magnitude stars were added or dropped. Many of the 
CCD observers recalibrated their measurements using the new sequence. 
However, since there were only a handful of CCD V-band observations of WZ\,Sge
prior to the 2001 superoutburst, this sequence change had a negligible effect on CCD V-band data prior to this outburst.
It could be, however, that individual observers started using the photometric standards sooner than this, especially at the faint end.  
In that time frame the visual estimates match the RoboScope measurements more closely.

We note that comparison stars fainter than V$\sim$13.5, typically, tend to be brighter when measured visually
(e.g., Bailey \&\ Howarth 1979, Zissel 2003). 
However, this cannot be the cause of the changes in discrepancy seen for WZ\,Sge, because the comparison stars used since about 1970 are more or less the same.
It merely applies to observations using older charts and sequences where no V-band calibrated comparison stars were used.

\begin{figure}[top]
\centering
  \includegraphics[height=.35\textheight,angle=-90]{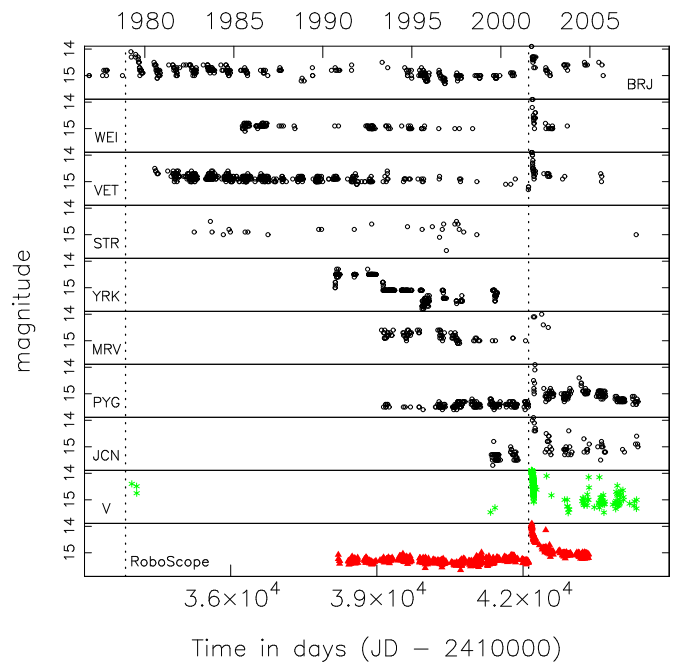}
  \caption{Visual measurements of individual AAVSO observers with more than
20 estimates during the two quiescent 1978--2001 and 2001--2007 time periods. 
The different observers are indicated with three-letter acronyms.
For reference we show the AAVSO V-band and RoboScope data at the bottom.
}
\label{individual}
\end{figure}

\subsection{Individually contributing observers}

In Fig.~\ref{individual} we show the measurements of eight individual amateur observers, together with the RoboScope data
and other V-band data as references. We only show data for those observers with more than 20 measurements in the 1978--2001 
quiescent time interval. 
For the first four observers (BRJ, WEI, VET, and STR, respectively; first four panels of Fig.~\ref{individual}), 
there is no evidence of any noticeable change in observing behaviour in the 1990's. 
Also the sixth (MRV) and seventh (PYG) observers do not show any change when comparing the pre-1997 and past-1997 measurements.
However, for the fifth observer (YRK; 
fifth panel of Fig.~\ref{individual}) measurements tend to be higher by up to about a magnitude in the beginning of the 1990's compared
to that at the end of the 1990's. The later observing behaviour may have an effect on the average visual
brightness estimates in the first part of the 1990's, and thus (partly) introduce a mismatch between the 
overall AAVSO dataset and RoboScope dataset. Excluding the data from this observer in the calculation of the 
average magnitudes, however, does not change the quantitative behaviour described in Sect.~\ref{discrepancy}.
In general, all the observers report a higher brightness after a superoutburst, with respect to the
brightness observed before a superoutburst. This is accords with what is seen in the RoboScope data.
Also, trends seen in the data for the first three, the fifth, and sixth observers (BRJ, WEI, VET, YRK, and MRV, respectively)
are consistent with a decline in brightness over the 1978--2001 quiescent interval. For the fourth (STR) and seventh
(PYG) observers this is not evident. However, the latter only has measurements from 1993, and the behaviour seen, as
well as brightness, is consistent with what is seen with the RoboScope.

Amateur CCD observations made during the mid- to late 1990's were usually unfiltered,
and mostly no report was made of the comparison star(s) used. In those early days of amateur CCD work, comparison stars were often chosen 
from sources that would not be used today. In quiescence, WZ\,Sge is close to the limit for visual observers;
towards this faint end, CCD magnitudes without making use of a filter have been reported to be 1--3 magnitudes fainter than visual 
estimates (e.g., Green 1997).
Also, because software for astronomers was scarce, data reduction was not as sophisticated as it is today. It may, therefore, be that some
of the magnitude estimates are affected in that time period. 
In the 1990's, CCD cameras became a mass-product so affordable for amateur astronomers. 
By the mid-1990's many amateurs therefore used CCDs for their observations.
However, this only affects the individual's measurements in an abrupt way, not for the whole ensemble, and certainly
not in any trend over a long time period. Moreover, only a few CCD observations of WZ\,Sge are available in that
time frame, so their impact is expected to be small.

During the 1990's many visual observers were switching from red-light illuminated paper charts 
to computer-generated charts on screens which emit significant blue light.
Using a red light to read charts will temporarily make the background sky appear bluer, while using a computer screen for charts 
will make the background sky appear redder. Visual estimates of luminous differences depend on contrast. 
A blue or a red star may thus appear fainter or brighter, respectively, than it was before, if one changed the way 
of using charts. When considering individual observers, the latter would imply a sudden change in their magnitude estimates. 
However, since WZ\,Sge's colour is bluer than red (see, e.g., Table 2 and Fig.~3), it would mean that the star's 
brightness based on visual estimates {\it increases} after changing to computer-generated charts, which is opposite the 
behaviour seen in the AAVSO and RoboScope datasets during the 1990's.

Changes in individuals equipment (e.g., use of a larger telescope) may also contribute to a sudden mismatch.
One of the AAVSO observers (BRJ), for example, had begun using a larger telescope in 1993 and 
experienced an approximate 0.5\,mag faintening of all his visual estimates made with this instrument.
However, judged from Fig.~\ref{individual}, the impact on the measurements of WZ\,Sge is less than 0.5\,mag.
Moreover, an individual's change in aperture (as well as other changes) cannot affect the trends seen in the overall measurements.
In particular, it will be difficult to understand then why the visual and V-band measurements do not match in the mid-90's, while they
do a couple of years later.

One may also be worried that some visual observations were (temporarily) contaminated by the nearby $\simeq$10$\arcsec$ star
(see Sect.~\ref{Stiening}). 
However, we regard it as unlikely that this star and WZ\,Sge could be confused 
by either visual observers or by automated software. When WZ\,Sge is at quiescence, the other star is somewhat brighter than WZ\,Sge
(see, e.g., Krzeminski \&\ Kraft 1964).
To the eye, the nearby star to WZ\,Sge, WZ\,Sge itself, and another nearby pair of similar brightness stars are very distinct.
We also checked a sample of the RoboScope data: the two stars were always detected separately.

\subsection{Response of the eye and the (Johnson) V-band filter}

Another effect that may have played a role is the difference in the eye's response with respect to that of a V-band filter.
Normally, the eye has a peak sensitivity at similar wavelengths (typically 555\,nm) as the
Johnson V-band ($\sim$550\,nm), but the response is considerably wider for the eye (see, e.g., Starr et al.\ 2006).
When a star is blue, more light is registered by the eye than with an instrument equipped with a Johnson V filter,
and thus a blue star may appear brighter in the visible with respect to the V-band.
In low-light conditions, i.e., at night, the situation is, however, somewhat different (so-called `Purkinje effect', see Purkinje 1825). 
With decreasing luminance both red and blue appear to darken. This is due to a narrowing of the wavelength range to which the 
eye responds as the luminance decreases to the point of peak sensitivity. With a further decrease in luminance, the eye's peak 
sensitivity moves toward the blue; i.e., it peaks around 500\,nm 
(see, e.g., Bowmaker \&\ Dartnall 1980). A blue object may again appear brighter or more intense than 
a red object at night (i.e., an observer's eye sees red objects as fainter than blue ones), and thus may introduce differences in the 
magnitude estimates by eye versus those with a V-band filter.

The multi-colour Stiening photometry of WZ\,Sge suggests that the U-band data showed some decline, whereas
the V and R band data may show a slight brightening in our period of interest (see Sect.~\ref{results});
i.e., the star was becoming intrinsically redder. If a star becomes redder over time,
estimates by eye may result in an apparent decrease in brightness in that time frame due to the Purkinje effect, 
and visual and V-band magnitudes may become more consistent with each other. 
Colour corrections have been introduced to better match visual and V-band observations, by making use of (B$-$V) measurements
(e.g., Bailey \&\ Howarth 1979, Stanton 1981, Collins 1999, Zissell 2003). However, in quiescence, WZ\,Sge shows B$-$V values 
during individual measurements between about 0 and $+$0.3, with an average value of about $+$0.12 (Krzeminski \&\ Smak 1971).
During a superoutburst, it is about $-$0.03, with individual measurements lying between $-$0.15 and $+$0.15
(Patterson 1978, Heiser \&\ Henry 1979, Howarth 1981, Patterson et al.\ 2002, Howell et al.\ 2004; see also Eskio\v glu 1963).
Using these values (and assuming similar colour values for the comparison stars) the colour correction is expected to be at 
most a tenth of magnitude, and can thus not explain the difference between the visual and V-band data.
Moreover, visual and V-band measurements after the 2001 superoutburst do not show
any discrepancy, where a similar intrinsic colour change is expected.

\subsection{Conclusion}

To conclude, there was no major observing sequence change prior to WZ\,Sge's 2001 superoutburst that would account for the general decline
from post-outburst to pre-outburst. Some sequence changes could have enabled observers to 
make more faint observations after 1973, which might have affected the trend of the average values in that time interval. 
However, the different brightness levels and the trends seen during the 1978--2001 quiescent cycle
are also evident in the other quiescent cycles (Sect.~3), strengthening it being intrinsic to the source.
There were some problems with individual observers in the 1990's, however, whose many data may affect the analysis of the
AAVSO light curve in that time frame. Individual effects may add systematic differences up to few tenths of a magnitude. However,
the largest discrepancy seen, $\simeq$0.7\,mag, we suspect to be a cumulative effect of various systematics, but we cannot 
pinpoint which effects were the dominating ones in the 1990--1997 time period.

\end{document}